\documentclass[useAMS]{mn2e}

\usepackage{graphicx,amssymb,amstext} 

\begin{document}

\title[PCA of the {\it Spitzer} IRS spectra of ULIRGs]{Principal component analysis of the {\it Spitzer} IRS spectra of ultraluminous infrared galaxies} 
\author[L. Wang et al.]{
\parbox[t]{\textwidth}{
Lingyu Wang$^1$\thanks{E-mail: lingyu.@sussex.ac.uk}, Duncan Farrah$^1$, Brian Connolly$^2$, Natalia Connolly$^3$, Vianney LeBouteiller$^4$, Seb Oliver$^1$, Henrik Spoon$^4$}
\\
$^1$Astronomy Centre, Department of Physics and Astronomy, University of Sussex, Falmer, Brighton BN1 9QH, UK\\
$^2$Department of Physics and Astronomy, University of Pennsylvania, Philadelphia, PA 19104-6396, USA\\
$^3$Physics Department, Hamilton College, Clinton, NY 13323, USA\\
$^4$Cornell University, Astronomy Department, Ithaca, NY 14853, USA\\
}

\date{Accepted . Received ; in original form }

\maketitle

\begin{abstract}
We present the first principal component analysis (PCA) applied to a sample of 119 {\em Spitzer} Infrared Spectrograph ({\em IRS}) spectra of local ultraluminous infrared galaxies (ULIRGs) at $z<0.35$. The purpose of this study is to objectively and uniquely characterise the local ULIRG population using all information contained in the observed spectra. We have derived the first three principal components (PCs) from the covariance matrix of our dataset which account for over $90\%$ of the variance. The first PC is characterised by dust temperatures and the geometry of the mix of source and dust. The second PC is a pure star formation component. The third PC represents an anti-correlation between star formation activity and a rising AGN. Using the first three PCs, we are able to accurately reconstruct most of the spectra in our sample. Our work shows that there are several factors that are important in characterising the ULIRG population, dust temperature, geometry, star formation intensity, AGN contribution, etc. We also make comparison between PCA and other diagnostics such as ratio of the 6.2 $\mu m$ PAH emission feature to the 9.7 $\mu m$ silicate absorption depth and other observables such as optical spectral type. An electronic version of the first three PCs of the local ULIRG population is available at http://astronomy.sussex.ac.uk/$\sim$lw94/PCA/.
\end{abstract}


\section{INTRODUCTION}

Ultraluminous infrared galaxies (ULIRGs), discovered from ground-based infrared photometry in the 1970s (Rieke \& Low 1972), are usually defined to be galaxies with bolometric luminosities from $8-1000$ $\mu m$ $>10^{12} L_{\odot}$. Over the last decade or so, we have learnt from optical and near-infrared imaging that ULIRGs are mostly interacting or merging systems (e.g. Armus, Heckman \& Miley 1987; Melnick \& Mirabel 1990; Hutchings \& Neff 1991; Clements et al. 1996; Murphy et al. 1996; Surace et al. 2000; Farrah et al. 2001; Veilleux, Kim \& Sanders 2002). From spectral energy distribution (SED) modelling, optical, UV and mid-infrared spectroscopic studies, X-ray and radio imaging, it seems that the power source in these galaxies is usually some combination of star formation and mass accretion onto the central black hole with the former being the dominant contributor and the latter usually increases its importance as a function of infrared luminosity (eg Rowan-Robinson \& Crawford 1989; Smith 1998; Farrah et al. 2003, 2007; Ptak et al. 2003; Franceschini et al. 2003; Klaas et al. 2001; Imanishi et al. 2003, 2007). A basic evolutionary picture is that two or possibly more gas-rich spiral galaxies collide with each other; The collision triggers bursts of star formation in centrally concentrated and compressed interstellar gas and active galactic nuclei (AGN) activity (Mihos \& Hernquist 1996; Sanders \& Mirabel 1996; Moorwood 1996; Lonsdale, Farrah \& Smith 2007). The merged galaxy then turns into an elliptical and perhaps eventually an optical QSO. 

With the advent of Infrared Space Observatory ({\em ISO}; Kessler et al. 1996) and Infrared Spectrograph ({\em IRS}; Houck et al. 2004) on {\em Spitzer} Space Telescope (Werner et al. 2004), mid-infrared spectroscopy has become a powerful tool in studying the nature of ULIRGs. Broad polycyclic aromatic hydrocardons (PAHs) features, the strongest of which are located at 6.2, 7.7, 8.6, 11.2 and 12.7 $\mu m$, are ubiquitous in normal galaxies and starburst systems with moderately intense UV radiation but weak or absent near an AGN. It indicates that the strength of PAHs is correlated with the origin of activity and thus can be used to separate a buried AGN from starburst. Another tool to disentangle the two energy sources is the optical depths of the silicate dust absorption features at 9.7 and 18 $\mu m$ which is associated with source geometry (Imanishi et al. 2007). High ionization fine structure lines also allow us to assess the dominant ionization mechanism. For example, the presence of the [NeV] line at 14.3 $\mu m$ requires photons of energy $>97$ eV and therefore indicates the presence of an AGN. Armus et al. (2007) found that ULIRGs selected from the IRAS Bright Galaxy Sample (Soifer et al. 1987) have a large range in spectral slope, silicate optical depths and PAH strengths indicative of a diverse and complex population. Using diagnostics such as the ratio of high- to low-excitation mid-infrared emission lines, the 7.7 $\mu m$ line-to-continuum ratio and the 6.2 $\mu m$ PAH emission and the 9.7 $\mu m$ silicate feature, most ULIRGs are found to be starburst-dominated with at least 50$\%$ have both starburst and AGN activity (eg Genzel et al. 1998; Rigopoulou et al. 1999; Lutz et al. 1999; Spoon et al. 2007).

However, there are some difficulties in using the aforementioned diagnostics, eg the mixing of the 7.7 $\mu m$ feature with the adjacent 8.6 $\mu m$ feature and the difficulty of accurately defining the underlying continuum, especially for strongly obscured sources. Different groups measure PAHs in different ways and sometimes their estimates can differ by a factor of two. In addition, different diagnostics often lead to different assessments of the dominant power source. Principal component analysis (PCA) has been used for spectral classification of optical galaxies with great success (e.g. Connolly et al. 1995; Bromley et al. 1998; Folkes et al. 1999; Madgwick et al. 2002, 2003; de Lapparent et al. 2003). In this paper, we present the first PCA of mid-infrared spectra which uses all information contained in the observed spectra while many traditional methods only use a few pixels of the whole spectrum and thus may lose important information. 

This paper is organised as follows. In Section 2, we briefly describe our sample of mid-infrared spectra. In Section 3, we introduce the PCA method, its implementation and the most important principal components (PCs) derived from the covariance matrix. The stability of the mean spectrum and the PCs is then investigated using a bootstrap resampling technique. In Section 4, we present eigenvector decomposition of a few selected spectra and discuss the effect of each PC. Our sample of spectra is then divided into eight types according to the sign of the contribution from each PC. Comparison with other diagnostics and other observables are presented in Section 5. Finally, conclusions and discussions are presented in Section 6.

\begin{table*}
\caption{Our sample of 119 nearby ULIGRs at redshift $z\le0.35$.}\label{table:thesample}
\begin{tabular}[pos]{llllllll}
\hline
Name            &  Redshift  & Name & Redshift & Name & Redshift & Name & Redshift\\
\hline
IRAS 00091-0738   & 0.12 & IRAS 06009-7716 & 0.12& IRAS 12514+1027 & 0.32& IRAS 19254-7245 & 0.06\\
IRAS 00183-7111   & 0.33 & IRAS 06035-7102 & 0.08& IRAS 13218+0552 & 0.20& IRAS 19297-0406 & 0.09\\
IRAS 00188-0856   & 0.13 & IRAS 06206-6315 & 0.09& IRAS 13335-2612S & 0.12& IRAS 19458+0944 & 0.10\\
IRAS 00199-7426   & 0.10 & IRAS 06361-6217 & 0.16& IRAS 13342+3932 & 0.18& IRAS 20037-1547 & 0.19\\
IRAS 00275-0044   & 0.24 & IRAS 07145-2914 & 0.01& IRAS 13352+6402 & 0.24& IRAS 20087-0308 & 0.11\\
IRAS 00275-2859   & 0.28 & IRAS 07598+6508 & 0.15& IRAS 13451+1232 & 0.12& IRAS 20100-4156 & 0.13\\
IRAS 00397-1312   & 0.26 & IRAS 08559+1053 & 0.15& IRAS 13509+0442 & 0.14& IRAS 20414-1651 & 0.09\\
IRAS 00406-3127   & 0.34 & IRAS 08572+3915 & 0.06& IRAS 13539+2920 & 0.11& IRAS 20551-4250 & 0.04\\
IRAS 00456-2904SW & 0.11 & IRAS 09039+0503 & 0.12& IRAS 14060+2919 & 0.12& IRAS 21208-0519N & 0.13\\
IRAS 01003-2238   & 0.12 & IRAS 09116+0334 & 0.15& IRAS 14070+0525 & 0.26& IRAS 21272+2514 & 0.15\\
IRAS 01166-0844SE & 0.12 & IRAS 09539+0857 & 0.13& IRAS 14252-1550 & 0.15& IRAS 23060+0505 & 0.17\\
IRAS 01199-2307   & 0.16 & IRAS 10091+4704 & 0.25& IRAS 14348-1447 & 0.08& IRAS 23128-5919 & 0.04\\
IRAS 01298-0744   & 0.14 & IRAS 10378+1109 & 0.14& IRAS 14378-3651 & 0.07& IRAS 23230-6926 & 0.11\\
IRAS 01355-1814   & 0.19 & IRAS 10485-1447W & 0.13& IRAS 15001+1433 & 0.16& IRAS 23253-5415 & 0.13\\
IRAS 01388-4618   & 0.09 & IRAS 10494+4424 & 0.09& IRAS 15206+3342 & 0.12& IRAS 23327+2913 & 0.11\\
IRAS 01494-1845   & 0.16 & IRAS 10565+2448 & 0.04& IRAS 15225+2350 & 0.14& IRAS 23498+2423 & 0.21\\
IRAS 01569-2939   & 0.14 & IRAS 11038+3217 & 0.13& IRAS 15250+3609 & 0.06& IRAS 23578-5307 & 0.12\\
IRAS 02054+0835   & 0.34 & IRAS 11095-0238 & 0.11& IRAS 15462-0450 & 0.10& IRAS 3C273 & 0.16\\
IRAS 02113-2937   & 0.19 & IRAS 11130-2659 & 0.14& IRAS 16300+1558 & 0.24& IRAS Arp220 & 0.02\\
IRAS 02455-2220   & 0.30 & IRAS 11223-1244 & 0.20& IRAS 16334+4630 & 0.19& IRAS Mrk1014 & 0.16\\
IRAS 02530+0211   & 0.03 & IRAS 11387+4116 & 0.15& IRAS 16468+5200E & 0.15& IRAS Mrk231 & 0.04\\
IRAS 03000-2719   & 0.22 & IRAS 11506+1331 & 0.13& IRAS 16487+5447 & 0.10& IRAS Mrk273 & 0.04\\
IRAS 03158+4227   & 0.13 & IRAS 11582+3020 & 0.22& IRAS 17028+5817 & 0.11& IRAS Mrk463 & 0.05\\
IRAS 03250+1606   & 0.13 & IRAS 12018+1941 & 0.17& IRAS 17044+6720 & 0.13& IRAS NGC6240 & 0.02\\
IRAS 03521+0028   & 0.15 & IRAS 12032+1707 & 0.22& IRAS 17068+4027 & 0.18& IRAS PG1119+120 & 0.05\\
IRAS 03538-6432   & 0.30 & IRAS 12071-0444 & 0.13& IRAS 17179+5444 & 0.15& IRAS PG1211+143 & 0.08\\
IRAS 04103-2838   & 0.12 & IRAS 12112+0305 & 0.07& IRAS 17208-0014 & 0.04& IRAS PG1351+640 & 0.09\\
IRAS 04114-5117   & 0.12 & IRAS 12127-1412NE & 0.13& IRAS 17463+5806 & 0.31& IRAS PG2130+099 & 0.06\\
IRAS 04313-1649   & 0.27 & IRAS 12205+3356 & 0.26& IRAS 18030+0705 & 0.15& IRAS UGC5101 & 0.04\\
IRAS 05189-2524   & 0.04 & IRAS 12359-0725 & 0.14& IRAS 18443+7433 & 0.13\\
\hline
\end{tabular}
\end{table*}

\section{THE DATA}
\label{sample}

\subsection{Sample Selection}

We selected our sample from two observing programs; those ULIRGs observed as part of the IRS Guaranteed Time program (Armus et al 2006; Spoon et al 2007; Farrah et al 2007; Desai et al 2007), and those observed by Imanishi et al (2007). Both the IRS-GTO and Imanishi samples were selected from the IRAS 1Jy and 2Jy surveys and together comprise virtually all known ULIRGs at $z\lesssim 0.5$. We imposed an upper redshift cut of $z=0.35$ to ensure we sample approximately the same wavelength range for each object, and removed a further eight objects as they have poor quality data in the longer wavelength IRS modules due to failed peak-ups or other observing difficulties. The resulting sample comprises 119 objects, listed in Table~\ref{table:thesample}.

\subsection{Observations}
All objects were observed with both orders of the Short-Low (SL) and Long-Low (LL) modules of the IRS. Observations were performed in staring mode. The targets were acquired by performing a high-accuracy IRS peak-up with the blue array on the target itself, or by peaking up on a nearby Two Micron All-Sky Survey (2MASS; Skrutskie et al. 2006) star and offsetting to the target. Each galaxy was observed at two nod positions within each of the IRS orders. The resulting spectra have a spectral resolution of R $\sim80$ over 5 – 38 $\mu$m.

\subsection{Data Reduction}
The data were processed through the {\it Spitzer} Science Center's pipeline software (version 18.7), which performs standard tasks such as ramp fitting and dark current subtraction, and produces Basic Calibrated Data (BCD) frames. Starting with these frames, we removed rogue pixels using the IRSCLEAN tool\footnote{The IRSCLEAN package can be downloaded from {\it http://ssc.spitzer.caltech.edu}} and campaign-based pixel masks. The individual frames at each nod position were then combined into a single image using the SMART software (Higdon et al. 2010). Sky background was removed from each image by subtracting the image for the same object taken with the other nod position (i.e. `nod-nod' sky subtraction). One-dimensional spectra were then extracted from the images using `optimal' extraction and default parameters (LeBouteiller et al. 2010). This procedure results in separate spectra for each nod and for each order. The spectra for each nod were inspected; features present in only one nod were treated as artefacts and removed. The two nod positions were then combined. The first and last 4 pixels on the edge of each order, corresponding to regions of decreased sensitivity on the array, were then removed, and the spectra in different orders merged, to give the final spectrum for each object.

\begin{figure}\centering
\includegraphics[height=4.0in,width=3.5in]{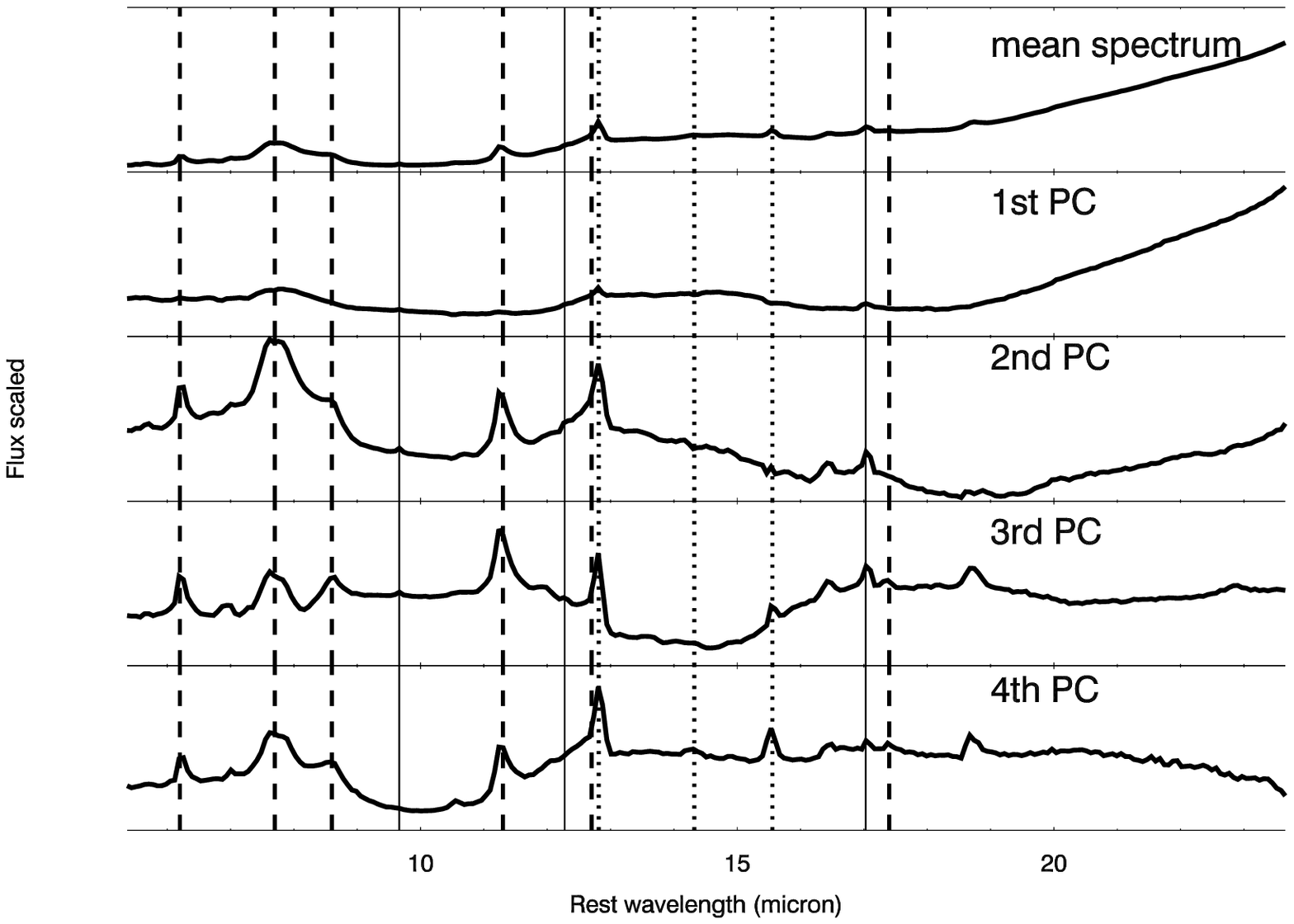}
\includegraphics[height=4.0in,width=3.5in]{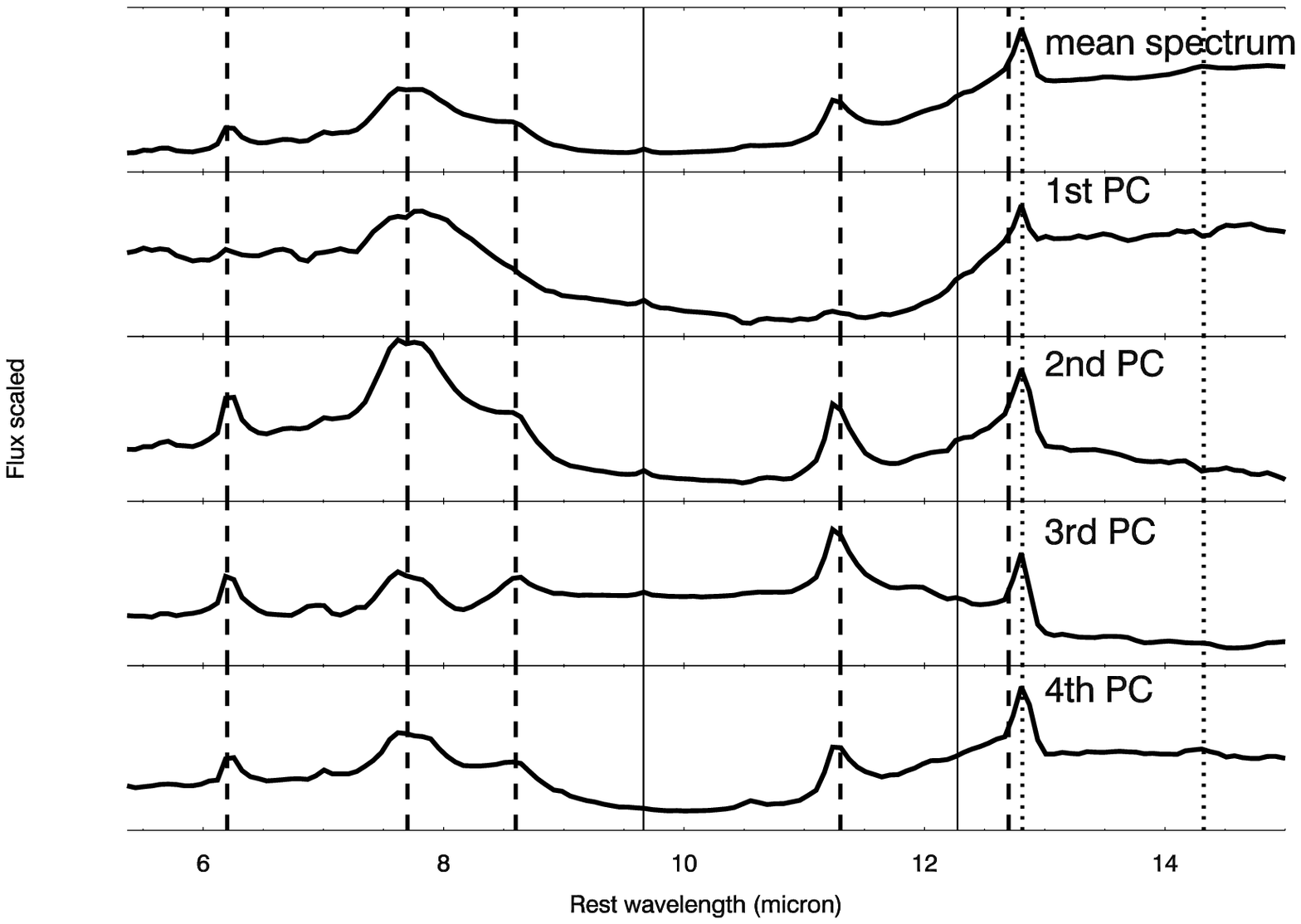}
\caption{Upper panel: The mean spectrum of all objects in our sample and the first four PCs. The dashed vertical lines mark the central location of the 6.2, 7.7, 8.6, 11.2 and 12.7 $\mu m$ PAH emissions. The thin vertical lines indicate the location of the molecular hydrogen lines at 9.66, 12.28 and 17.03 $\mu m$. The dotted vertical lines indicate the positions of the neon fine-structure lines, [Ne II] 12.8, [Ne v] 14.3 and [Ne III] 15.6 $\mu m$. We have chosen to show PCs with PAH features in emission. The sign of each PC is arbitrary. Lower panel: A closer look at the 5 - 15 $\mu m$ region of the mean spectrum and the four PCs.}
\label{fig:meanspectrum}
\end{figure}

\begin{figure}\centering
\includegraphics[height=4.0in,width=3.5in]{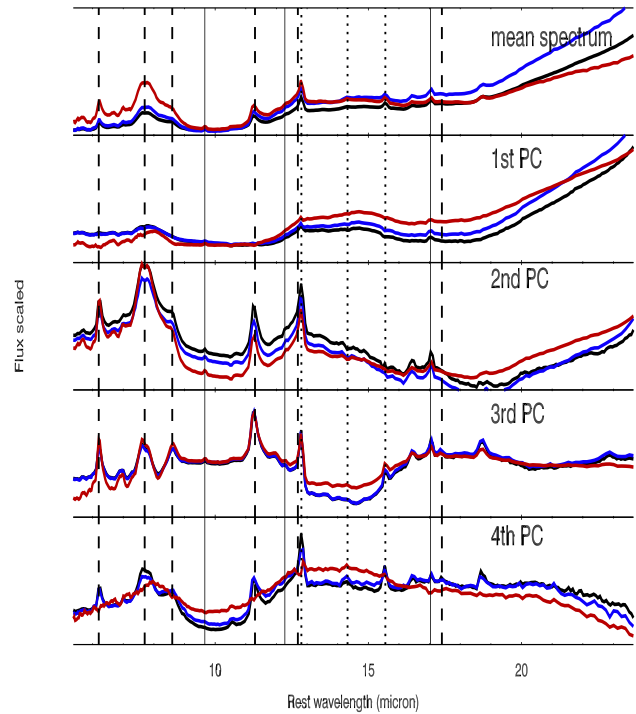}
\caption{The mean / median spectrum of all objects in our sample and the first four PCs derived using different set-ups of the analysis. The black curves correspond to the mean spectrum and the first four PCs under (S1) (see text for definitions of different set-ups). The blue curves correspond to the median spectrum and the derived PCs under (S2). The red curves correspond to the median spectrum and the derived PCs under (S3).}
\label{fig:ComparisonPCs}
\end{figure}

\begin{figure}\centering
\includegraphics[height=4.0in,width=3.5in]{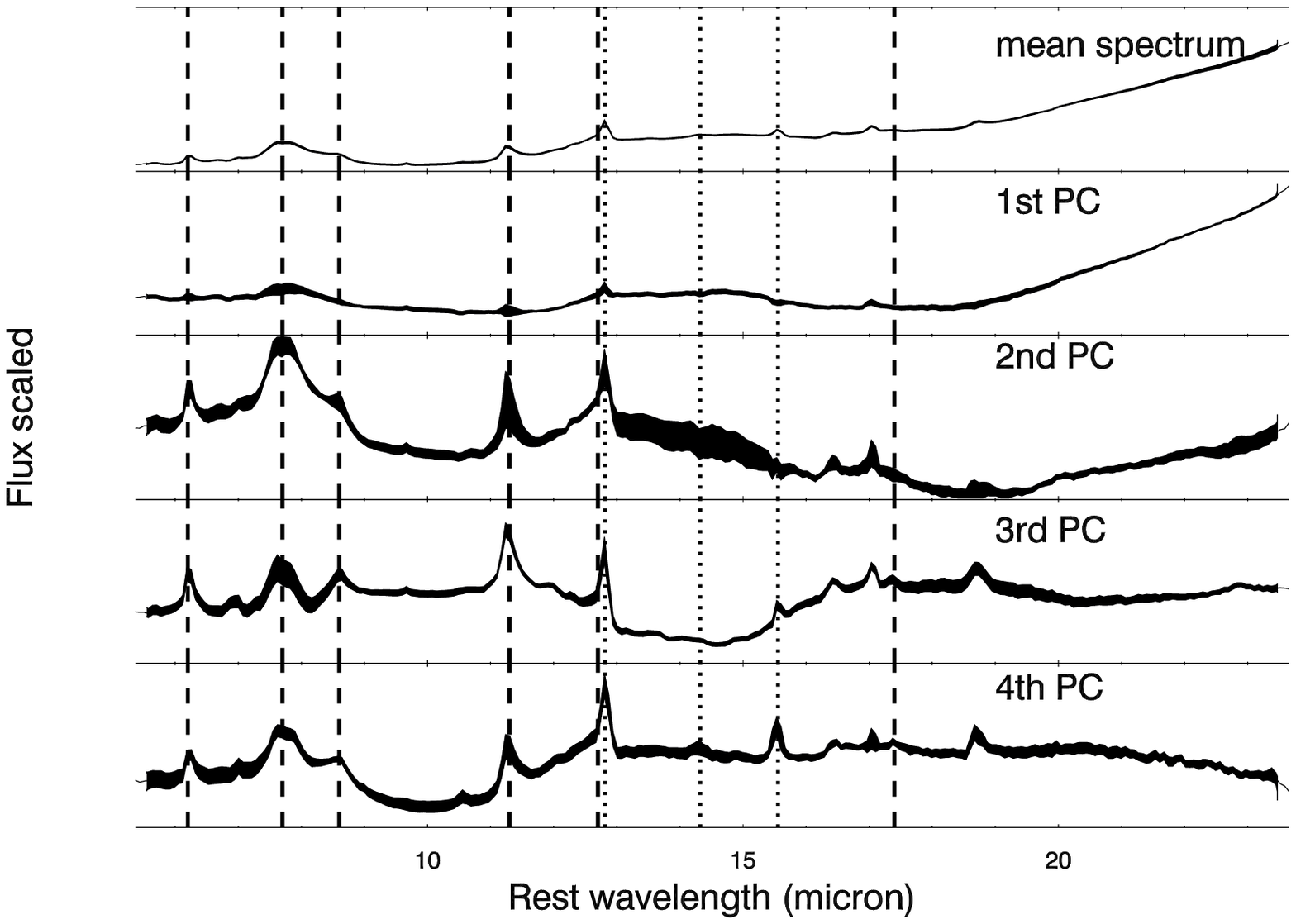}
\caption{The $1-\sigma$ uncertainty range in the mean spectrum and the first four PCs (from top to bottom) from 100 bootstrap realisations of the original sample. The mean spectrum stays more or less the same.}
\label{fig:BSPpcs}
\end{figure}

\section{Principal component analysis (PCA)}

\subsection{A brief introduction of PCA}

PCA is a non-parametric tool, often used to reduce a complex dataset to a simpler structure and searching for correlations. A primary benefit of PCA analysis arises from quantifying the importance of each dimension for describing the variability of a data set. 

Suppose we have $M$ spectra. The spectrum of the $i$th galaxy, a sequence of $N$ numbers ($f_1^i, \cdots,f_N^i$), can be treated as a vector in an $N$-dimensional space, where $N$ is the number of spectral channels. We can then form a covariance matrix of all the spectra,
\begin{equation}
C_{ij} = \frac{1}{M} \Sigma_{k=1}^{k=M} f_k^i f_k^j, 1 \le i, j \le N.
\end{equation}
The covariance matrix can be diagonalized by the matrix of its orthogonal eigenvectors (or eigenspectrum) $E = [e_1, e_2, \cdots, e_N]$,
\begin{equation}
C =  E D E^T,
\end{equation}
where $D$ is a diagonal matrix and its diagonal terms are the eigenvalues of the corresponding eigenvectors. In the language of PCA, the eigenvectors, each of which is a linear combination of the original spectra, are principal components (PCs). In addition, PCs are ordered by their importance, ie the percentage of variance they account for. In some cases, only a few PCs are needed to describe the data which is why PCA is used to reduce the dimensionality of a complex dataset. The underlying assumption of PCA is that eigenvectors associated with large variance reveal important structures, while those associated with small eigenvalues represent noise. However, PCA does have limitations because it assumes that the variance is sufficient to represent the data. Therefore, PCA only applies to multivariate Gaussian distributions.

\subsection{Implementation of PCA and the derived PCs}
First of all, we need to de-redshift the observed spectrum to its rest frame. We have chosen a rest-frame wavelength range from 5.37 - 23.7 $\mu m$ which is covered by all spectra in our dataset and then divided it into 268 equally-spaced bins in linear wavelength space. Each spectrum is normalised so that the mean flux over the whole wavelength range is unity. The mean spectrum of our sample is then subtracted off from each spectrum. We form a covariance matrix of all mean-subtracted spectra and then derive the eigenvectors/PCs. The mean spectrum and the first four PCs are shown in Fig.~\ref{fig:meanspectrum}. We emphasise that the PCs represent the difference between the observed spectra and the mean spectrum of the sample.

In the mean spectrum, we can clearly see features such as broad PAH emissions, neon fine-structure lines and broad absorption from amorphous silicate at 9.7 and 18 $\mu m$. The first component PC1 is characterised by weak PAH emissions, weak neon lines, fairly strong silicate absorption and a steep spectral slope. In both PC2 and PC3, strong PAHs are the dominant features. Although PC4 is shown in here, we do not interpret PC4 as a proper PC because the distribution of our sample in the PC1 - PC4 plane is non-Gaussian. For now, PC4 is interpreted as extra features in the spectra of quasars (more discussion in Section 5.5). 

One caveat to bear in mind is that the derived PCs might be different in different set-ups of the analysis. In addition to the set-up described above (referred to as (S1) hereafter)  we consider two more scenarios: (S2) Instead of re-sampling the spectra to an array with equal width in $\lambda$ and subtracting the mean spectrum off all spectra, we re-sample the spectra to an array with equal width in $\log \lambda$ and subtract the median spectrum off. (S3) Same as S2 except that the spectra are now changed to $\nu F_\nu$ rather than $F_\nu$. The different set-ups essentially give different effective weighting to different parts of the spectra. In Fig.~\ref{fig:ComparisonPCs}, we compare the mean / median spectrum of our sample and the first four PCs derived under different set-ups. We find that although there are noticeable changes, the qualitative features in the mean / median spectrum and the derived PCs remain the same. For example, PC1 is characterised by weak PAH features, strong silicate absorption and steep spectral slope under different set-ups. We also note that there are almost no line features in PC4 under (S3). However, given that PC4 is not a proper PC and it is not needed to reconstruct the spectra for the majority of our objects, we conclude that the differences in the derived mean / median spectrum and the major PCs are immaterial to our results presented in the following sections.

\subsection{A stability study: a bootstrap approach}
We use a bootstrap method\footnote{Each bootstrap realisation is generated by sampling the original dataset with replacement 119 times. As a result, some objects might be sampled more than once and some object in the original dataset might be absent in a particular realisation.} on the original dataset to study the stability of the mean spectrum and the PCs. We repeat the steps in Section 3.2 on each bootstrap realisation. Fig.~\ref{fig:BSPpcs} shows the $1-\sigma$ region of the mean spectrum and the first four PCs using 100 realisations. The mean spectrum is very stable. In PC1, the stable features are the steep spectral slope and broad silicate absorption. The former implies a large amount of cold dust and the latter a large dust column density. PAH features vary from weak to nearly absent suggesting that PAHs are not important. If there is a negative PC1 contribution, we will get weaker silicate absorption (or even silicate emission if it is sufficiently negative) and a flatter/bluer spectrum but it has almost nothing to do with PAH features. In PC2, we generally see strong PAH emissions and a flat overall spectral shape. If there is a negative PC2 contribution, then PAH features will be weakened but the spectral shape will not be altered much. In PC3, again we see a flat overall spectrum and strong PAHs. There is a broad trough centred at $\sim14\mu m$. If PC3 is negative, then we will see a strong mid-infrared continuum indicating the presence of hot dust.

\subsection{Eigenvector decomposition}
\begin{figure}\centering
\includegraphics[height=3.3in,width=3.5in]{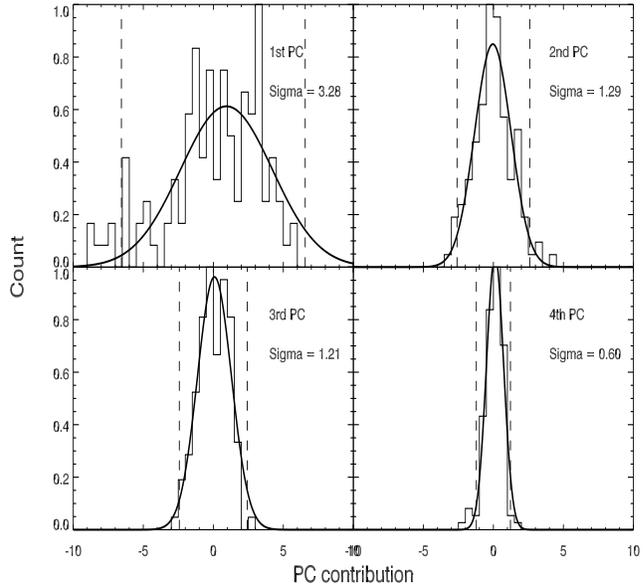}
\caption{Histogram of contributions to each object in our sample from each of the first four PCs, normalised so that the peak of the distribution is equal to one. The heavy curves are Gaussian fits to the histograms. The standard deviation of each Gaussian fit is shown in each panel. The dashed lines mark the $2-\sigma$ range. A greater width of the Gaussian distribution indicates that more sources need a contribution from that particular PC.}
\label{fig:hist_PC}
\end{figure}

We can decompose every spectrum in our sample by projecting it onto the first four PCs,
\begin{eqnarray}
f & = &\textrm{Mean~Spectrum} + C1 \times PC1 + C2 \times PC2 \nonumber \\
  & ~ & + C3 \times PC3 + C4 \times PC4,
\end{eqnarray}
where $C1, C2, C3$ and $C4$ are the coordinates along PC1, PC2, PC3 and PC4 respectively. The histogram of contributions from each PC is plotted in Fig.~\ref{fig:hist_PC}. In each panel, the distribution can be fitted by a Gaussian function, the width of which decreases from PC1 to PC4 as expected. The width of PC4 is significantly narrower than the first three PCs meaning that the number of objects which need significant contributions from PC4 is small. Indeed, only about six objects need a large contribution from PC4. Four of these objects show silicate emission, one show deep silicate absorption and no PAH emission and one show strong [NeV] indicating that PC4 is mainly an indicator of powerful AGNs. Indeed, in Fig.~1 and Fig.~2, we can see strong neon lines and strong silicate absorption when PC4 is positive and silicate emission when it is sufficiently negative.

\begin{figure*}\centering
\includegraphics[height=4.3in,width=7.3in]{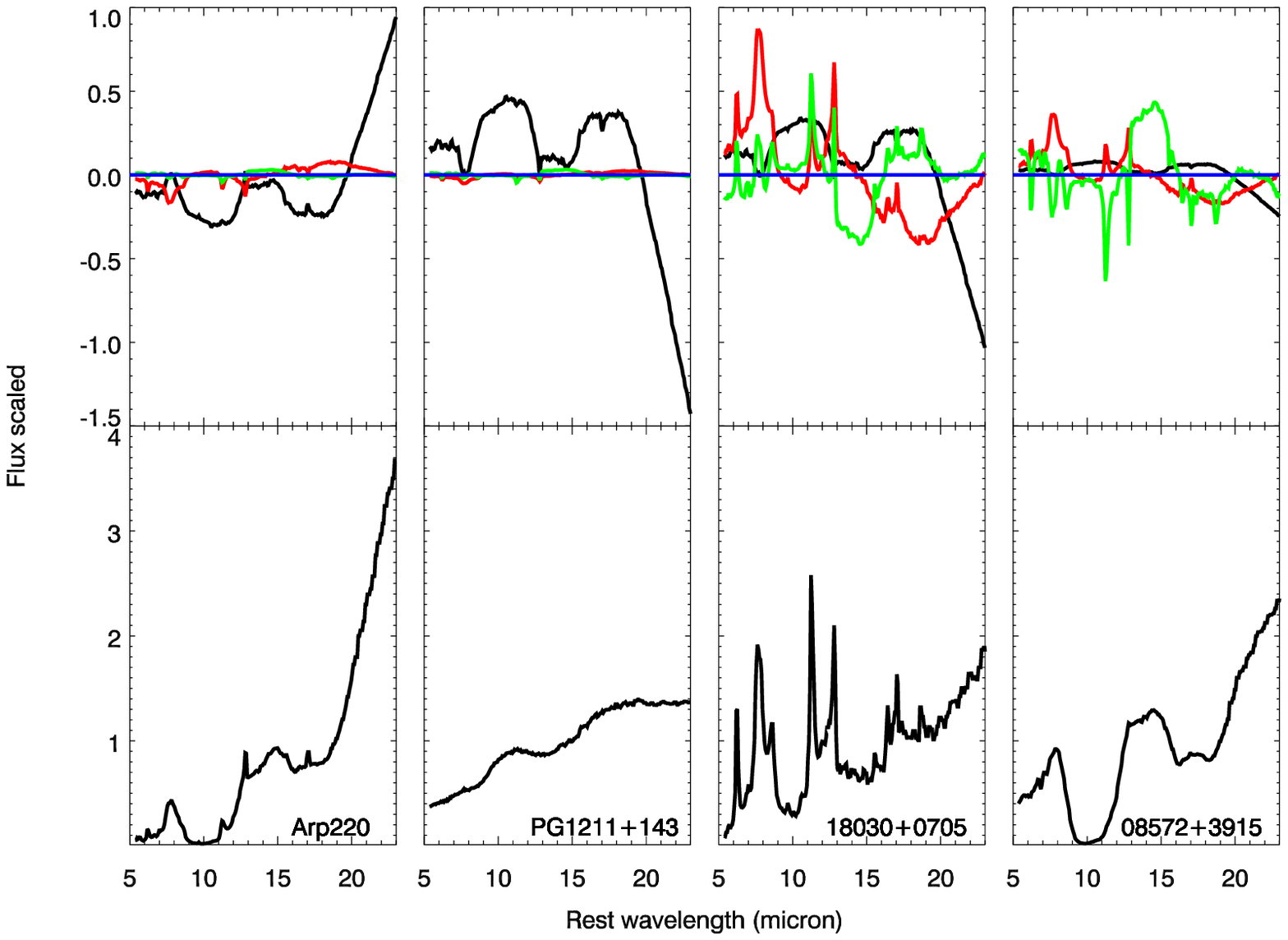}
\includegraphics[height=4.3in,width=7.3in]{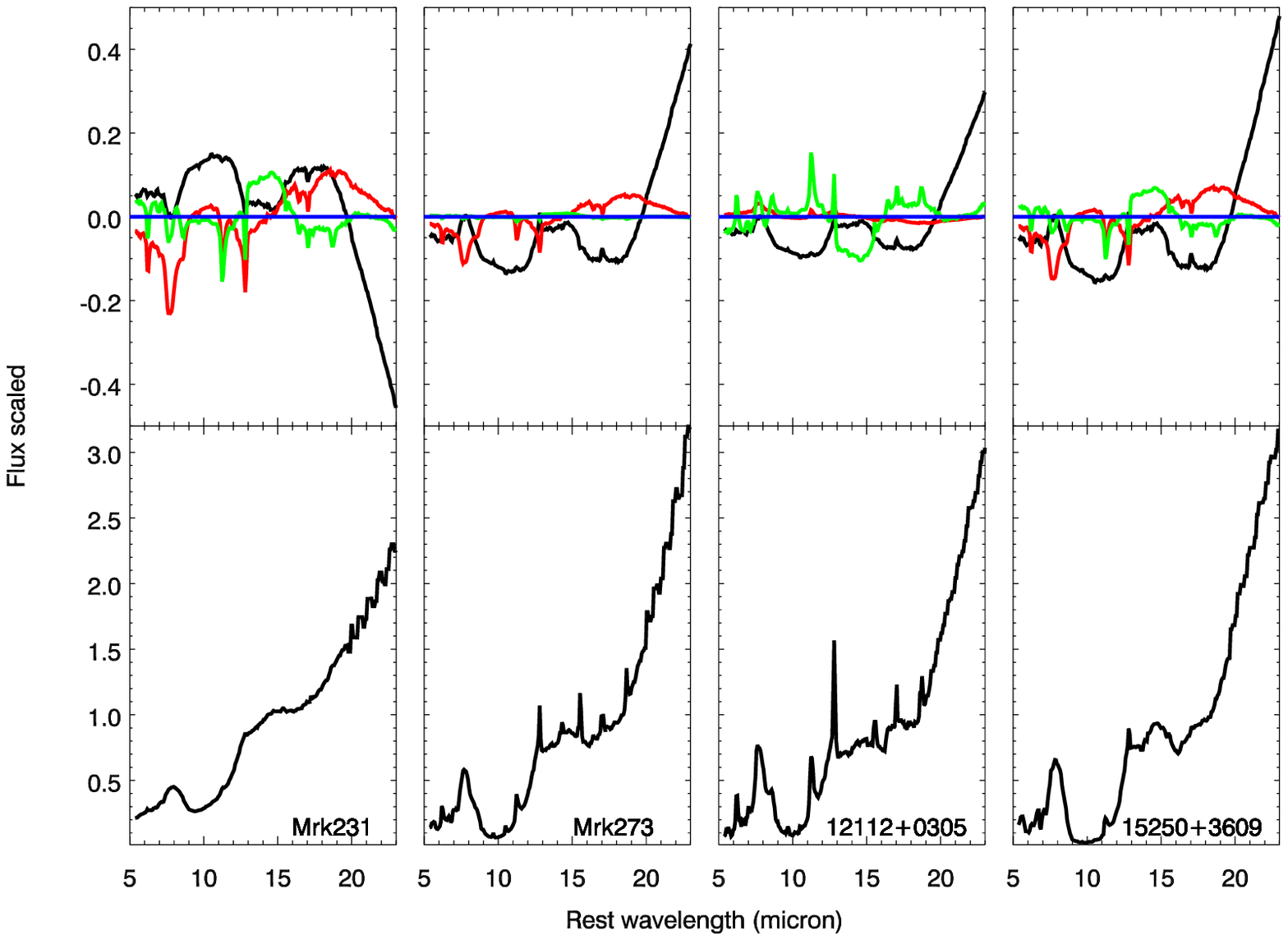}
\caption{In each panel, the lower solid curve is the original spectrum and the four curves above it show the contributions from the first three PCs (black - PC1; red - PC2; green - PC3). The blue line represents the zero level.}
\label{fig:casestudy2}
\end{figure*}

\section{What does each PC do?}
\subsection{Case studies using eigenvector decomposition}

Fig.~\ref{fig:casestudy2} show spectral decomposition of eight ULIRGs, six of which are from the IRAS BGS sample. The starburst-dominated Arp 220 has a huge amount of cold dust around 30 K. It has the largest positive PC1 resulting in the reddest spectrum and very strong silicate absorption. PG 1211 is a quasar and it has the largest negative PC1. It confirms that PC1 roughly indicates silicate absorption depth. F18030 has the largest positive PC2 and the largest positive PC3, consistent with the fact that F18030 has the strongest PAH emissions in all galaxies in our sample. F08572, optically classified as LINER, has the largest negative PC3 to attenuate PAH emission. It has the bluest spectrum and very deep silicate absorption.  Mrk 231 has a large negative PC1 implying the presence of hot dust. It also has a large negative contribution from both PC2 and PC3 resulting in very weak PAHs and strong mid-infrared continuum. All of these suggest than it is AGN-dominated which is in agreement with its optically classification as Seyfert 1 (broad-line AGN). The spectrum of Mrk 273, mainly characterised by a large positive PC1 and small amount of negative PC2, suggests that it is a less obscured version of Arp 220.

\subsection{Spectral classification using PCs}

\begin{figure}\centering
\includegraphics[height=2.8in,width=3.4in]{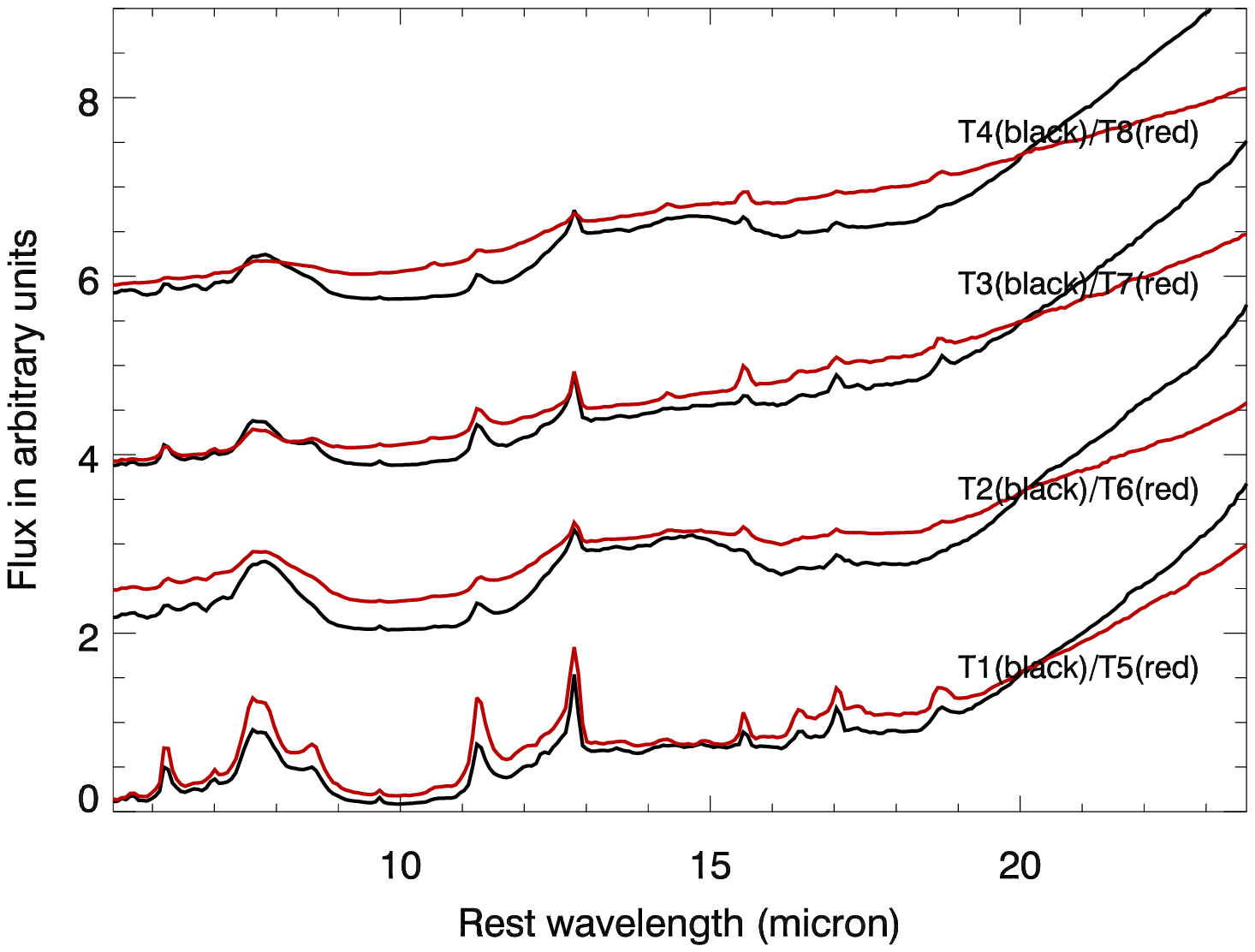}
\includegraphics[height=2.8in,width=3.4in]{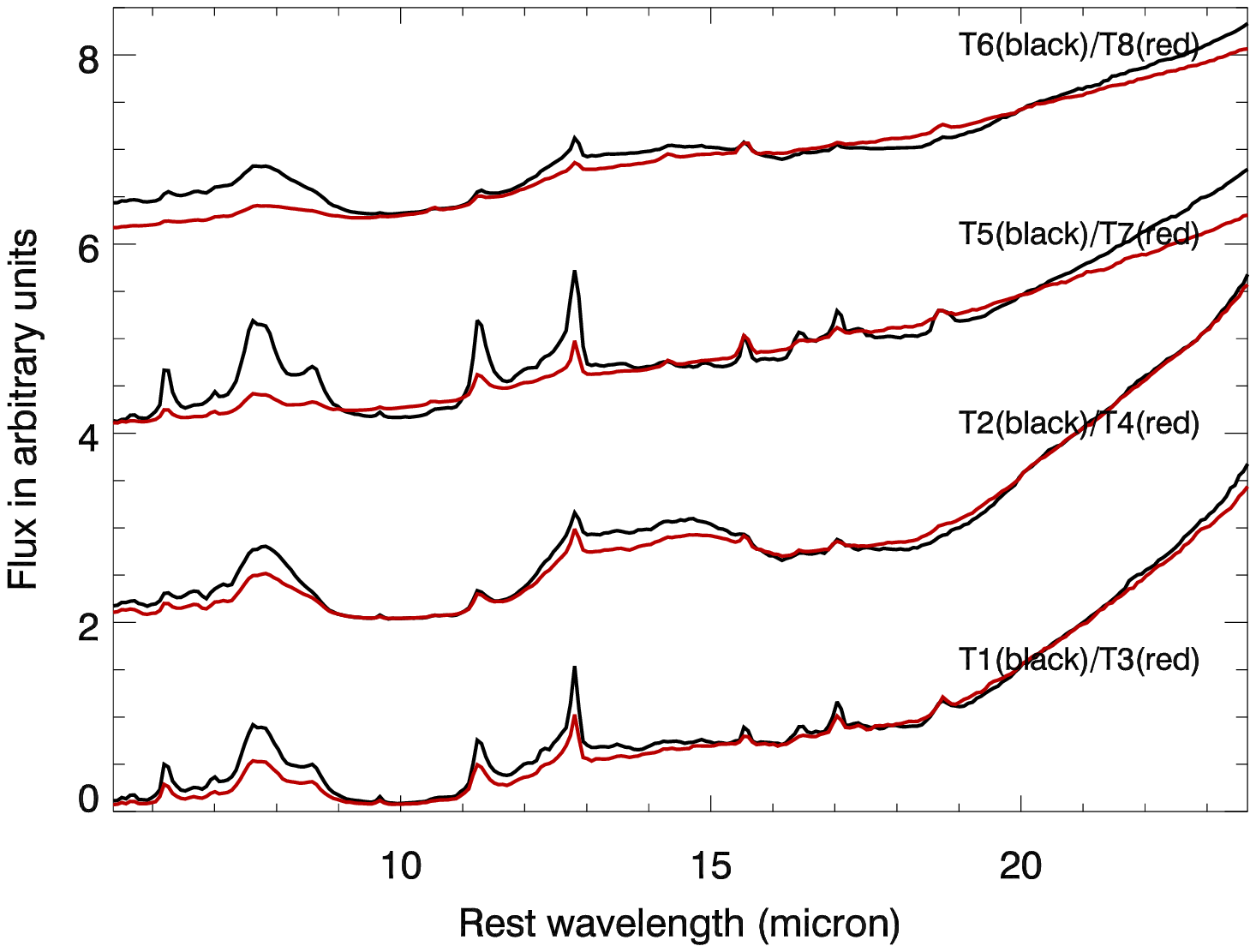}
\includegraphics[height=2.8in,width=3.4in]{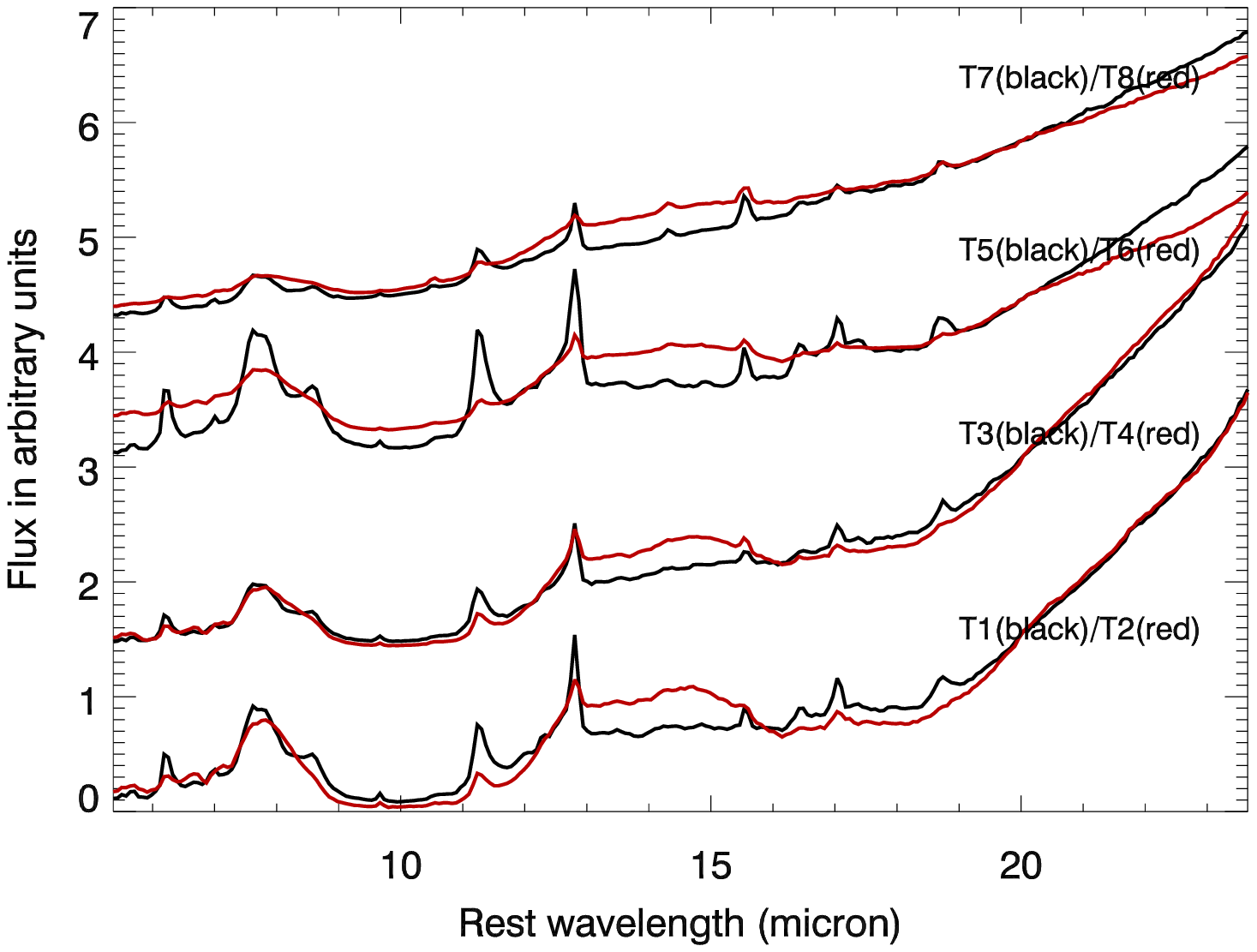}
\caption{Top: The mean spectra of the eight types, shown in four groups so that the only difference in each group is the sign of PC1 (black line - $PC1>0$; red line - $PC1<0$). Adjacent spectra are normalised at 20 $\mu m$. Middle: The only difference in each group is the sign of PC2 (black line - $PC2>0$; red line - $PC2<0$). Bottom: The only difference in each group is the sign of PC3 (black line - $PC3>0$; red line - $PC3<0$).}
\label{fig:MeanSpectrum_NewClasses1}
\end{figure}

In Section 4.1, we have looked at the effect of PCs on individual galaxies. But what is the general effect of each PC? Given that only the first three PCs are needed to reconstruct the majority of our spectra, it seems natural that we divide our sample into eight types according to sign of the contribution from each of the first three PC. In Table~\ref{table:Subclasses}, we have listed the definitions of the eight types and the number of objects in each type. 

In the top panel in Fig.~\ref{fig:MeanSpectrum_NewClasses1}, we have grouped the mean spectra of the eight types into four groups each of which contains two types. The only difference between the two spectra in each group is the sign of PC1. For example, the pair at the bottom of Fig.~\ref{fig:MeanSpectrum_NewClasses1} shows the mean spectra of objects in T1 and T5. Both types have positive PC2 and positive PC3 but opposite signs of PC1. We can see that while PAH features stay more or less the same in each pair, silicate absorption depth as well as the continuum shape changes. Similarly, in Fig.~\ref{fig:MeanSpectrum_NewClasses1} we have grouped the mean spectra into four groups with the only difference in each group being the sign of PC2. In each group, the red curve has weaker PAH emissions than the black curve while silicate absorption and the continuum shape remain more or less unchanged. Lastly, in the bottom panel in Fig.~\ref{fig:MeanSpectrum_NewClasses1}, the only difference in the mean spectra in each group is the sign of PC3. Galaxy types with PC3$<0$ have weaker PAH emission and stronger silicate absorption.  

\begin{table}
\caption{Eight types of ULIRG spectra.}\label{table:Subclasses}
\begin{tabular}[pos]{lll}
\hline
Type            & Definition & Number of objects \\
\hline
T1    &    PC1 $>$ 0, PC2 $>$ 0, PC3 $>$ 0 & 17\\
T2    &    PC1 $>$ 0, PC2 $>$ 0, PC3 $<$ 0 & 16\\
T3    &    PC1 $>$ 0, PC2 $<$ 0, PC3 $>$ 0 & 13\\
T4    &    PC1 $>$ 0, PC2 $<$ 0, PC3 $<$ 0 & 17\\
T5    &    PC1 $<$ 0, PC2 $>$ 0, PC3 $>$ 0 & 14\\
T6    &    PC1 $<$ 0, PC2 $>$ 0, PC3 $<$ 0 & 11\\
T7    &    PC1 $<$ 0, PC2 $<$ 0, PC3 $>$ 0 & 15\\
T8    &    PC1 $<$ 0, PC2 $<$ 0, PC3 $<$ 0 & 16\\
\hline
\end{tabular}
\end{table}

\begin{figure}\centering
\includegraphics[height=3.0in,width=3.5in]{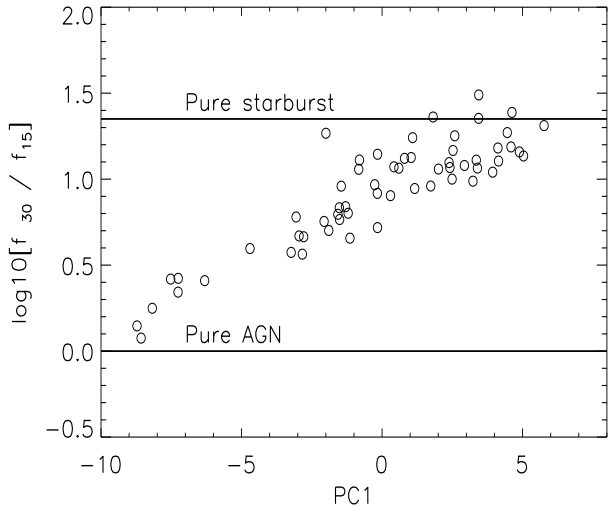}
\caption{The contribution of PC1 versus the $f_{30}/f_{15}$ continuum ratio. The two horizontal lines with $\log10( f_{30} / f_{15}) = 0$ and 1.35 are the zero points for pure AGN and starburst, respectively.}
\label{fig:QUEST}
\end{figure}

\begin{figure}\centering
\includegraphics[height=3.0in,width=3.5in]{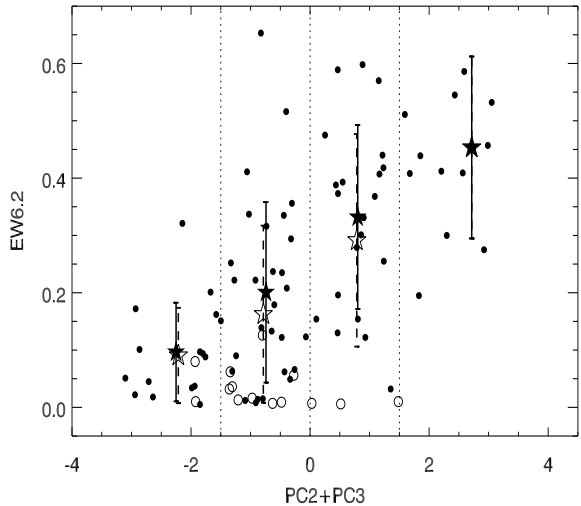}
\caption{The total contribution of PC2 and PC3 versus the PAH $6.2 \mu$m equivalent width (EW$6.2$) for all objects in our sample (open circles - objects with optical spectral type classified as Seyfert 1; small filled circles - everything else). We divide our objects into four bins of the total contribution of PC2 and PC3, which are indicated by the vertical dotted lines. The empty stars represent the median values of the EW$6.2$ for all galaxies in each bin of (PC2 + PC3) and the filled stars represent the median EW6.2 excluding all Seyfert 1 type galaxies.}
\label{fig:EW62} 
\end{figure}

\begin{figure}\centering
\includegraphics[height=4.0in,width=3.5in]{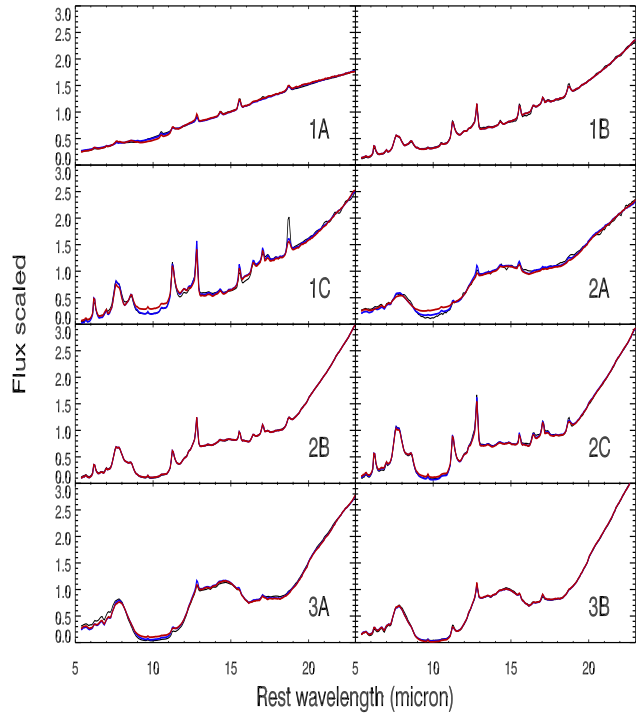}
\caption{The average spectra for eight classes defined in Spoon et al. (2006) and the reconstructed spectrum using the first three (red line) and the first four (blue line) principal components. The black curve represents the original spectra of the eight classes.}
\label{fig:SpoonClass}
\end{figure}

\section{Comparison with other diagnostics}

\subsection{Comparison with the $f_{30}/f_{15}$ continuum ratio}

Veilleux et al. (2009) presented results from {\it Spitzer} {\it IRS} observations of 74 ULIRGs and 34 Palomar Green (PG) quasars in the local Universe ($z<0.3$). They find that the $f_{30}/f_{15}$ continuum ratio can be used as a surrogate of the PAH-free, silicate-free MIR/FIR ratio to search for AGN activity. There are 59 sources in their analysis which are also included in our sample. In Fig.~\ref{fig:QUEST}, we compare the contributions of PC1 in these sources with the $f_{30}/f_{15}$ continuum ratio. A good correlation between PC1 contribution and the $f_{30}/f_{15}$ continuum ratio can be seen in the plot. ULIRGs near the zero point for pure AGN ($\log10( f_{30} / f_{15}) = 0$) have large negative PC1 values. On the other hand, ULIRGs near the zero point for pure starburst ($\log10( f_{30} / f_{15}) = 0$) have large positive PC1 values. The scatter in this correlation seems to increase as PC1 contribution increases which could indicate that PC1 contribution does not correspond directly to AGN contribution in objects where the dominant power source is star formation.

\subsection{Comparison with PAH 6.2 $\mu$m equivalent width}

Another commonly used diagnostic for star formation activity is the PAH 6.2 $\mu$m equivalent width (EW6.2). In Fig.~\ref{fig:EW62}, we compare the total contribution of PC2 and PC3 with EW6.2 for all galaxies in our sample. In general, objects with large EW6.2 have large positive (PC2+PC3) contribution and objects with negligible EW6.2 have large negative (PC2+PC3) contribution. A few galaxies with optical spectral type classified as Seyfert 1 have near zero PAH emission at 6.2 $\mu$m but positive (PC2+PC3) contribution. This is because in these objects, there are large negative contributions from PC1 which results in negative PAH features. So, a positive total contribution from PC2 and PC3 is used to cancel out these negative PAH features.

\subsection{Comparison with the Spoon IRS diagnostic diagram}
\begin{figure*}\centering
\includegraphics[height=4.0in,width=7.0in]{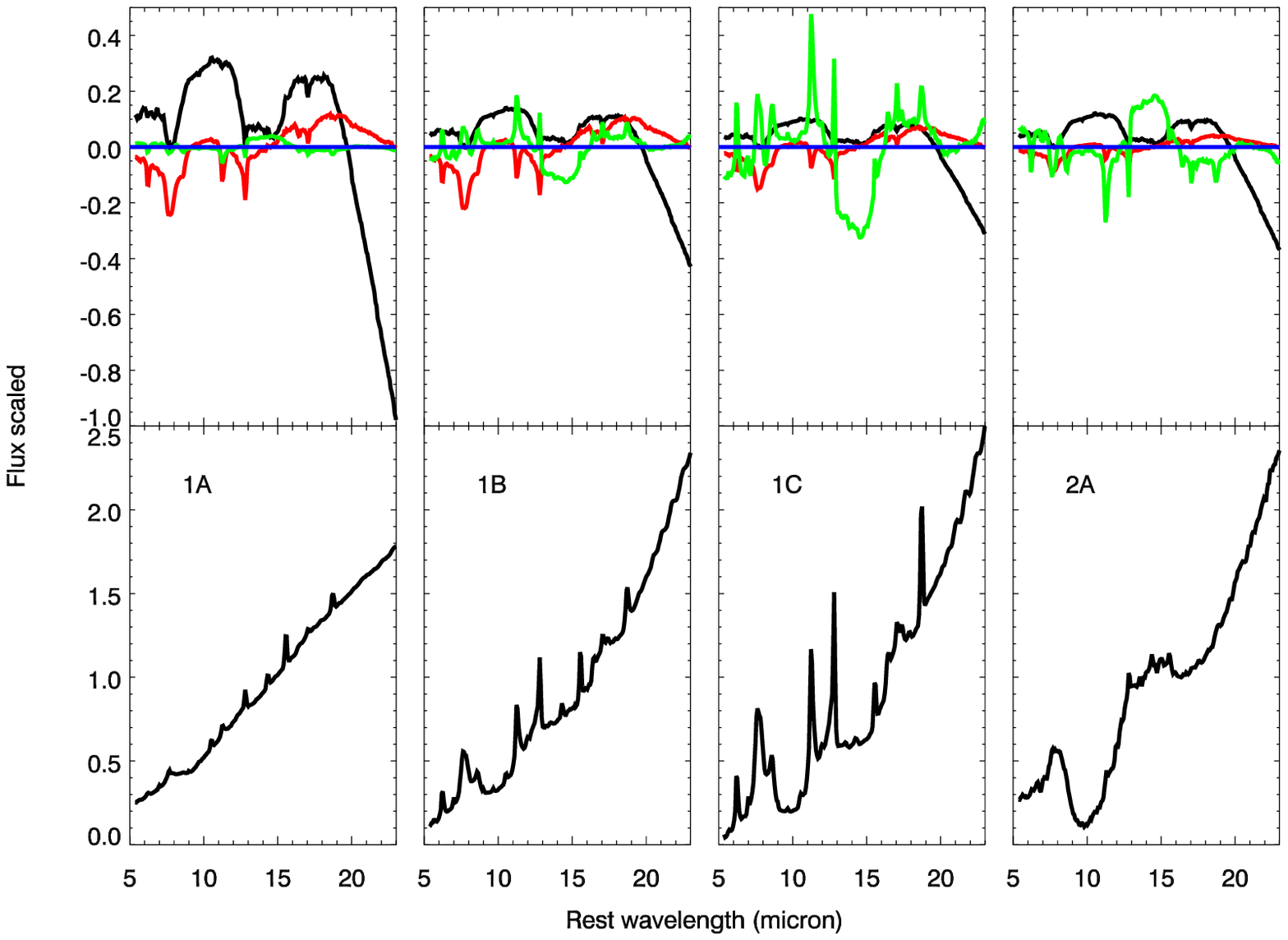}
\includegraphics[height=4.0in,width=7.0in]{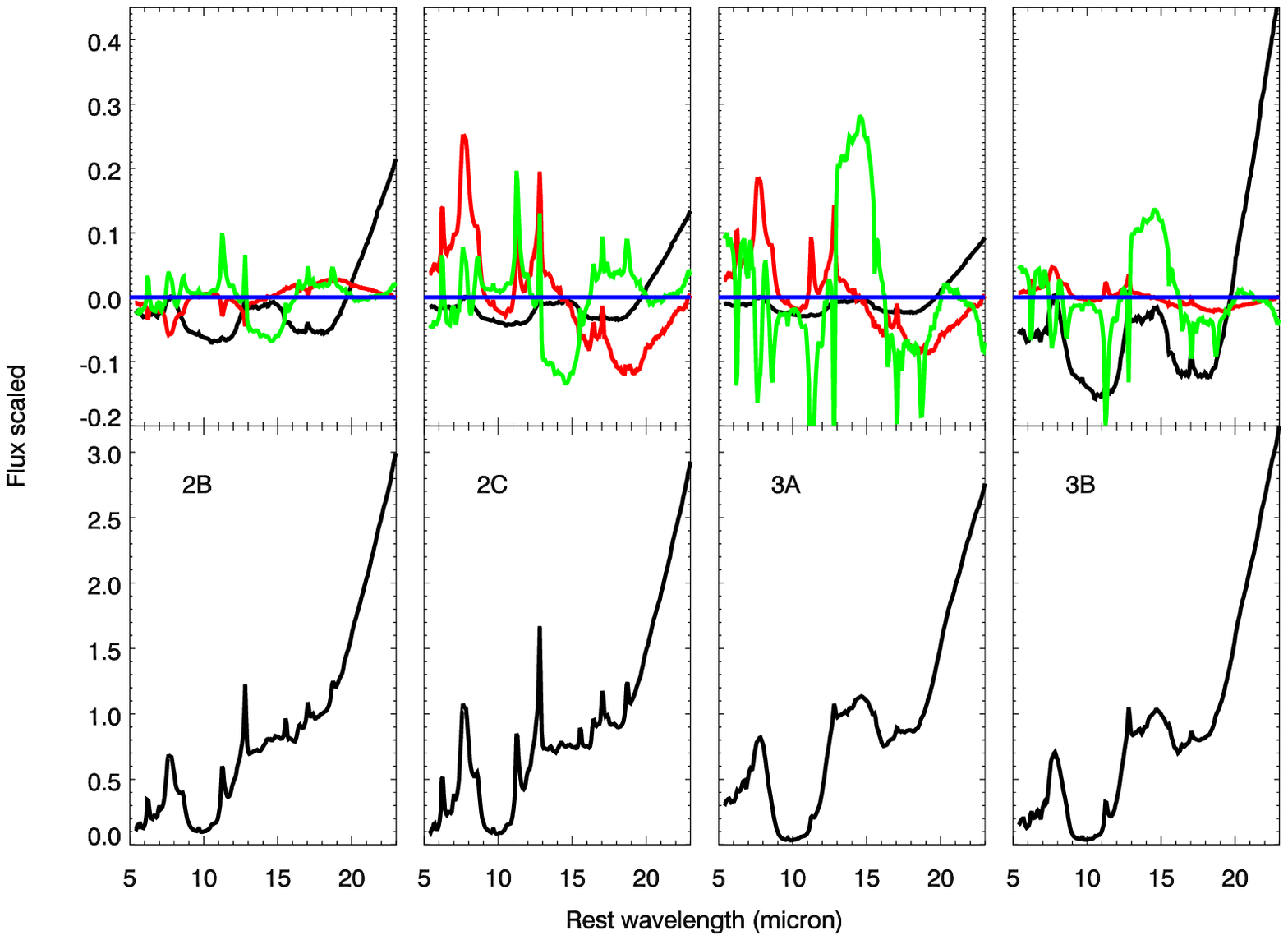}
\caption{The average spectrum of each of the classes (the lower curve in each panel) and the contribution of each of the first three PCs (black - PC1; red - PC2; green - PC3). The blue line represents the zero level.}
\label{fig:Class_reconstruction2}
\end{figure*}

 Spoon et al. (2007) used the equivalent width (EW) of the 6.2 $\mu m$ PAH feature and the optical depth of the 9.7 $\mu m$ silicate feature as a new diagnostic tool to divide galaxies into various classes ranging from continuum-dominated AGN hot dust spectra and PAH-dominated starburst spectra to absorption-dominated spectra of deeply obscured galactic nuclei. Fig.~\ref{fig:SpoonClass} shows the mean spectra of the eight classes defined in Spoon et al. (2007) and the reconstructed spectra using the first three or four PCs. Clearly, the first three PCs already provide an accurate reconstruction for all classes. Fig.~\ref{fig:Class_reconstruction2} shows the contribution of each PC to each class. 

The class 1A spectrum is a nearly featureless hot dust continuum with very weak silicate absorption centred at $9.7~\mu m$. It is mainly composed of a large negative PC1 and a large negative PC2. The former results in weak silicate absorption and a blue spectrum and the latter causes weak PAHs. The class 1B spectrum clearly shows PAH emission in addition to a hot dust continuum. Its spectral decomposition is similar to that of the class 1A, however, the class 1B also requires a large positive PC3 to show strong PAH features.

In the class 1C, we see strong PAH emissions caused by large positive PC3. The class 2C spectrum is similar to class 1C apart from stronger silicate absorption and steeper 20-30 $\mu m$ continuum. Accordingly, the class 2C has a large positive PC2, positive PC3 and positive PC1.

The class 2B spectrum has weaker PAHs than class 2C. It has a large positive PC1, positive PC3 and negative PC2. The class 3B has stronger silicate absorption than 2B. It has a positive PC1 and negative PC3. A negative PC3 contributes most to the class 3A which has the maximum silicate absorption. It confirms that a negative contribution from PC3 both reduces PAH emission and deepens silicate absorption. The class 2A, dominated by a large negative PC3 and PC1, has weaker silicate absorption than 3A. 

\begin{figure}\centering
\includegraphics[height=2.8in,width=3.4in]{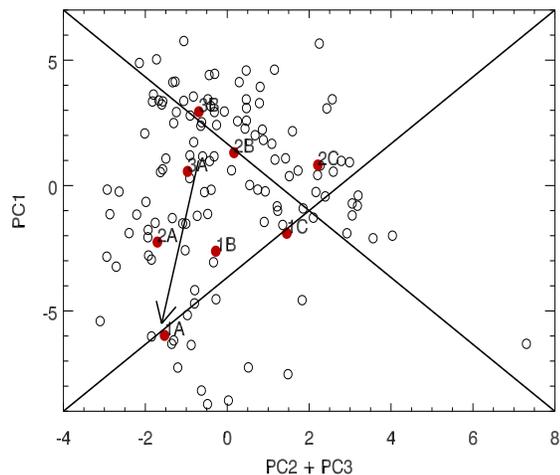}
\caption{PC1 versus the sum of PC2 and PC3. The red dots represent Spoon classes and the open circles represent our sample. The diagonal lines and the arrow (to represent the effect of continuum dilution) are used to indicate the similarities between this plot and the fork diagram presented in Spoon et al. (2007).}
\label{fig:PC_distribution2}
\end{figure}

 In Spoon et al. (2007), galaxies are found to distribute along two branches in the plane defined by the 6.2 $\mu m$ PAH EW and the 9.7 $\mu m$ silicate absorption strength. For galaxies with strong star formation, we expect to see a positive total contribution from PC2 and PC3. On the other hand, for galaxies with weak or no PAH features, the total contribution from PC2 and PC3 should be negative. In other words, the sum of PC2 and PC3 can serve as an approximate measure of the intensity of star formation activity. The contribution from PC1 corresponds to the strength of silicate absorption. In Fig.~\ref{fig:PC_distribution2}, we plot the sum of PC2 and PC3 against PC1 where a pattern similar to the fork diagram presented in Spoon et al. (2007) is seen. The horizontal branch in the fork diagram in Spoon et al. (2007), from 1A, 1B to 1C with increasing PAH equivalent width at 6.2 $\mu m$ roughly corresponds to the diagonal line going from the bottom left corner to the top right corner. Similarly, the diagonal branch in the fork diagram, from 3A, 2B to 1C with decreasing silicate absorption strength and increasing PAH equivalent width at 6.2 $\mu m$, seems to roughly correspond to the diagonal line going from the top left corner to the bottom right corner. We emphasise that Fig.~\ref{fig:PC_distribution2} does not show a complete fork diagram. To see the horizontal fork (1A-1B-1C) in full glory, the non-ULIRG AGN sample needs to be included as well. We also note that there are more transition sources showing up in Fig.~\ref{fig:PC_distribution2} compared to the fork diagram. This is because we have included sources from other Spitzer ULIRG programs besides the GTO sources used in Spoon et al. (2007).

\subsection{AGN contribution to bolometric luminosity}
\begin{figure}\centering
\includegraphics[height=3.in,width=3.4in]{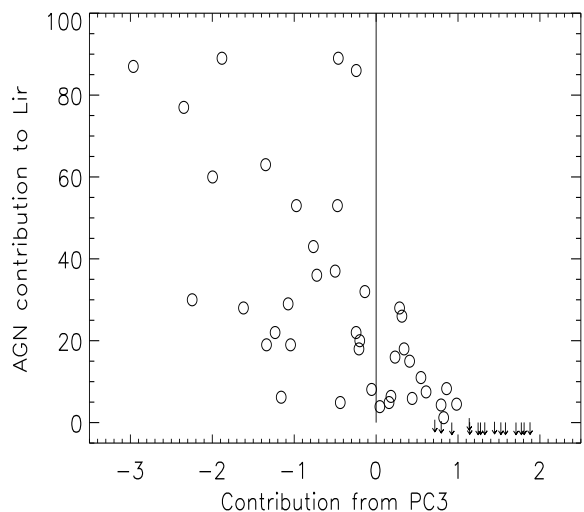}
\caption{PC3 contributions versus AGN contribution to the bolometric luminosity. Arrows indicate sources with upper limit estimates of its AGN contribution.} 
\label{fig:Nardini}
\end{figure}

Nardini et al. (2008; 2009) analysed the contribution AGN and starbursts to the bolometric luminosity based on a 5-8 $\mu m$ region of the spectra. They found evidence for AGN activity for $\sim$70\% of their sample which consists of 71 ULIRGs at $z<0.15$. 55 objects in their sample are found in our dataset. The AGN emission is modelled as a featureless power law $f_{\lambda} \propto \lambda^{1.5}$ with an exponential attenuation $e^{-\tau(\lambda)}$, where the optical depth is supposed to follow $\tau(\lambda) \propto \lambda^{-1.75}$. The SB template is built upon five brightest pure starbursts. In the left panel of Fig.~\ref{fig:Nardini}, we have plotted the contribution from PC3 as a function of Nardini et al.'s estimate of the AGN contribution to the bolometric infrared luminosity. We can see a clear correlation between contributions from PC3 and $\alpha_{\textrm{bol}}$. It supports the idea that a negative contribution from PC3 turns on AGN activity as discussed in the above sections.

\subsection{Optical spectral type}
\begin{figure}\centering
\includegraphics[height=3.in,width=3.4in]{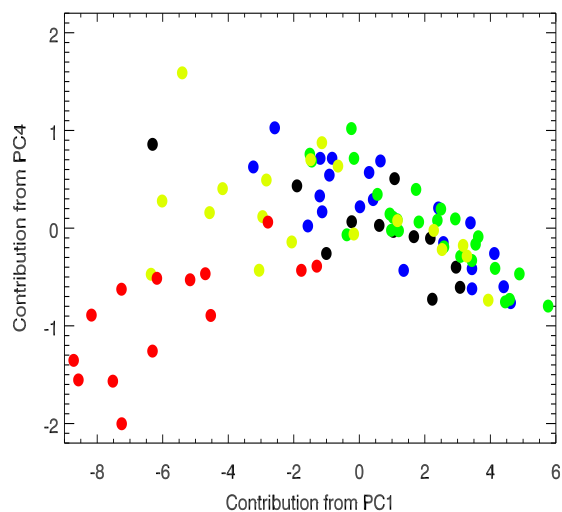}
\caption{Contributions from PC4 versus that from PC1, colour-coded by optical classes (blue dots - HII; green dots - LINER; yellow dots - Seyfert 2; red dots - Seyfert 1; black dots - unknown optical classes). }
\label{fig:OpticalClasses}
\end{figure}

Fig.~\ref{fig:OpticalClasses} shows the distribution of objects in the PC1-PC4 plane, colour-coded by their optical classifications including HII, LINER, Seyfert 1 and Seyfert 2. Clearly, this distribution is non-Gaussian which means `PC4' can not be interpreted as a proper PC. However, we can say that the few spectra which need large PC4 contributions are outside the parameter space spanned by the first three PCs. The most obvious feature in Fig.~\ref{fig:OpticalClasses} is that almost all Seyfert 1 type galaxies have a large negative PC4 contribution while objects of other types have a mean zero PC4 contribution with a small scatter. It shows again that PC4 has a remarkably good correspondence with optical QSOs.

\section{Discussion and conclusion}\label{discussions and conclusions}
This is the first time to our knowledge that principal component analysis has been applied to mid-infrared spectra. It is a simple yet powerful analysis tool. The first principal component PC1 mainly constrains the dust temperature and the geometry of the distribution of source and dust. Both the second and the third principal component, PC2 and PC3, seem to regulate the intensity of star formation activity. In addition, a large negative contribution from the third principal component corresponds to a brightening AGN. For a few sources with spectral features indicating a dominant AGN (e.g. silicate emission, strong [NeV]), the first three principal components are not enough in order to accurately reproduce the observed spectra and a fourth principal component is required.

Using principal component analysis, we are effectively finding an orthogonal basis to describe the variance which is assumed to fully characterise the objects under study. So the question is what mechanism/physics is causing spectra to vary. Are spectral differences related to different evolutionary stages of the ULIRG population? If so, by adding a positive or negative contribution from each PC to adjust the relative strength of various features (e.g. PAH emission, silicate absorption, spectral slope), the evolutionary stages of ULIRGs are modified. In this paper, we have compared our principal component analysis with various other diagnostics (such as the $f_{30}/f_{15}$ continuum ratio, the PAH 6.2 $\mu$m equivalent width and 9.7 $\mu$m silicate absorption strength). There are some tentative evidence that the principal components are linked to the evolutionary stages. The effect of each PC on the evolution of the ULIRG population will be further investigated using radiative transfer models in a future paper.

\section*{ACKNOWLEDGEMENTS}
We thank the referee for helpful comments which led to improvements in this paper. L. Wang is supported by UK's Science and Technology Facilities Council grant ST/F002858/1. D. Farrah is supported by UK's Science and Technology Facilities Council Advanced Fellowship. This work is based on observations made with the {\it Spitzer Space Telescope}, which is operated by the Jet Propulsion Laboratory, California Institute of Technology under a contract with NASA.


\begin{thebibliography}{99}
\bibitem{aa} Armus L., Heckman T.M., Miely G.K., 1987, AJ, 94, 831
\bibitem{bb} Armus L., 2007, ApJ, 656, 148
\bibitem{vv} Bromley G.C., Press W.H., Lin H., Kirshner R.P., 1998, ApJ, 505, 25
\bibitem{ff} Clements D.L., Sutherland W.J., McMahon R.G., Saunders W., 1996, MNRAS, 279, 477
\bibitem{rr} Connolly A.J., Szalay A.S., Bershady M.A., Kinney A.L., Calzetti D., 1995, AJ, 110, 1071
\bibitem{tete} de Lapparent V., Galaz G., Bardelli S., Arnouts S., 2003, A\&A, 404, 831 
\bibitem{dbdf} Desai V. et al., 2007, ApJ, 669, 810
\bibitem{xcvb} Farrah D. et al., 2001, MNRAS, 326, 1333
\bibitem{en} Farrah D., Afonso J., Efstathiou A., Rowan-Robinson M., Fox M., Clements D., 2003, MNRAS, 343, 585
\bibitem{tmndgh} Farrah D. et al., 2007, ApJ, 667, 149
\bibitem{cvbn} Franceschini A. et al., 2003, MNRAS, 343, 1181 
\bibitem{1} Folkes S. et al., 1999, MNRAS, 308, 459
\bibitem{2} Genzel R. et al. 1998, ApJ, 498, 579
\bibitem{3} Higdon et al., 2004, PASP 116, 975
\bibitem{4} Houck J.R. et al., 2004, ApJS, 154, 18
\bibitem{5} Hutchings J.B., Neff S.G., 1991, AJ, 101, 434
\bibitem{6} Imanishi M., Terashima Y., Anabuki N., Nakagawa T., 2003, ApJ, 596, L167
\bibitem{7} Imanishi M., Dudley C.C., Maiolino R., Maloney P.R., Nakagawa T., Risaliti G., 2007, ApJS, 171, 72
\bibitem{8} Kessler M.F. et al., 1996, A\&A, 315, L27
\bibitem{9} Klaas U. et al., 2001, A\&A, 379, 823
\bibitem{10} LeBouteiller V., Bernard-Salas J., Sloan G.C., Barry D.J., 2010, PASP, 122, 321L
\bibitem{11} Lonsdale C.J., Farrah D., Smith H.E., 2006, in Astrophysics Update 2., ed. J. W. Mason (Heidelberg: Springer), 285
\bibitem{12} Lutz D. Veilleux S. Genzel R., 1999, ApJ, 517, L13
\bibitem{13} Madgwick D.S. et al., 2002, MNRAS, 333, 133
\bibitem{14} Madgwick D.S. et al., 2003, ApJ, 599, 997
\bibitem{15} Melnick J., Mirabel I.F., 1990, A\&A, 231, 19
\bibitem{16} Mihos J.C., Hernquist L., 1996, ApJ, 464, 641
\bibitem{17} Moorwood A.F.M., 1996, SSRv, 77, 303
\bibitem{18} Murphy T.W. Jr., Armus L., Matthews K., Soifer B.T., Mazzarella J.M., Shupe D.L., Strauss M.A., Neugebauer G., 1996, AJ, 111, 1025
\bibitem{19} Nardini E., Risaliti G., Salvati M., Sani E., Imanishi M., Marconi A., Maiolino R., 2008, MNRAS, 385, L130
\bibitem{20} Nardini E., Risaliti G., Salvati M., Sani E., Watabe Y., Marconi A., Maiolino R., 2009, MNRAS, 399, 1373
\bibitem{21} Ptak A., Heckman T., Levenson N.A., Weaver K., Strickland D., 2003, ApJ, 592, 782
\bibitem{22} Rieke G.H., Low F.J., 1972, ApJ, 176, L95
\bibitem{23} Rigopoulou D., Spoon H.W.W., Genzel R., Lutz D., Moorwood A.F.M., Tran Q.D., 1999, AJ, 118, 2625
\bibitem{24} Rodriguez Zaurin J., Tadhunter C.N., Gonzalez Delgado R.M., 2009, MNRAS, 400, 1139
\bibitem{25} Rowan-Robinson M., Crawford, 1989, MNRAS, 238, 523
\bibitem{26} Sanders D.B., Mirabel I.F., 1996, ARA\&A, 34, 749
\bibitem{27} Skrutskie et al., 2006, AJ, 131, 1163 
\bibitem{28} Soifer B.T. et al. 1987, ApJ, 320, 238
\bibitem{29} Spoon H.W.W., Marshall J.A., Houck J.R., Elitzur M., Hao L., Armus L., Brandl B.R., Charmandaris V., 2007, ApJ, 654, L49 
\bibitem{30} Surace J.A., Sanders D.B., Evans A.S., 2000, ApJ, 529, 170
\bibitem{31} Veilleux S., Kim D.-C., Sanders D.B., 2002, ApJ, 143, 315
\bibitem{32} Veilleux S. et al., 2009, ApJS, 182, 628
\bibitem{33} Verma A., Charmandaris V., Klaas U., Lutz D., Haas M., 2005, Space Science Reviews, 119, 355
\bibitem{34} Werner M.W. et al., 2004, ApJS, 154, 1
\end{thebibliography}
\end{document}